\documentclass[journal]{IEEEtran}
\usepackage{url}
\usepackage{subfigure}

\usepackage[dvips]{graphicx}
\DeclareGraphicsExtensions{.eps}

\usepackage[cmex10]{amsmath}
\usepackage{array}

\begin{document}

\title{Synthesis-by-analysis of BCH Codes}

\author{\IEEEauthorblockN{Atta Ul Mustafa and Ghulam Murtaza}\\
\IEEEauthorblockA{Department of Electrical Engineering, FAST, Islamabad, Pakistan\\
National University of Sciences and Technology, Islamabad, Pakistan\\
atta.mustafa@nu.edu.pk, azarmurtaza@hotmail.com}
}

\maketitle

\begin{abstract}

In this paper we propose a technique to blindly synthesize the generator polynomial of BCH codes. The proposed technique involves finding Greatest Common Divisor (GCD) among different codewords and block lengths. Based on this combinatorial GCD calculation, correlation values are found. For a valid block length, the iterative GCD calculation results either into generator polynomial or some of its higher order multiples. These higher order polynomials are factorized under modulo-2 operation, and one of the resulting factors is always the generator polynomial which further increases the correlation value. The resulting correlation plot for different polynomials shows very high values for correct block length and valid generator polynomial. Knowing the valid block length and generator polynomial, all other parameters including number of parity-check digits $(n-k)$, minimum distance $d_{min}$ and error correcting capability $t$ are readily exposed.\\
\end{abstract}


\section{Introduction} \label{sec:introduction}
Error control coding is mandatory to combat unavoidable random and burst errors in digital communication channel. There exist various error control codes amongst which Bose-Chaudhuri-Hocquenghem (BCH) cyclic codes are very famous and widely used in digital communication channels. These codes are characterized by block length $n$, number of parity-check digits $(n-k)$ and minimum distance $d_{min}$.The generator polynomial of BCH codes is specified as Least Common Multiple (LCM) of minimal polynomials $\phi_i(X)$ where $1\leq i\leq 2t$, $t$ being error correcting capability of the code.

In a problem of eavesdropping a communication channel, no prior knowledge is available except the eavesdropped bitstream. The source information is packed into a number of different layers before sending it to the communication channel. In this scenario, one has to blindly estimate different parameters at each layer. Very few papers deal with the problem of synthesis and reconstruction of error control codes from eavesdropped bitstreams. Rice \cite{Rice95} presented a technique to estimate the parameters of rate $1/n$ convolutional code which was generalized by Filiol \cite{Filiol97} for other rates as well as for punctured convolutional codes. Burel \cite{Burel2003} suggested blind estimation of encoder and interleaver characteristics based on linear algebra theory. Barbier \cite{Barbier2005} analyzed different techniques to blindly recover the parameters of turbo-code encoder. In 2006, Cluzeau \cite{Cluzeau2006} introduced a version of Gallager algorithm with weighted parity-check equations to recover LDPC and other block codes.

In this paper, synthesis-by-analysis of BCH codes is presented. The proposed technique focuses on the parameter estimation at channel coding layer in general and on BCH codes in particular. In our work, the key parameter to be estimated is the generator polynomial for a valid block length $n$. Knowing the valid block length $n$ and generator polynomial $g(X)$, all other parameters can be readily found and BCH codes can be decoded without any prior knowledge of the transmission side.

We assume that we have access to the eavesdropped BCH encoded bitstream. This assumption is simulated by generating test vectors for a range of BCH codes $(n,k,t)$. For a specific $(n,k,t)$ code, the test vectors are passed to the proposed algorithm and GCD is found for two codewords in first iteration. The algorithm then steps through different available codewords in a combinatorial manner. For each combination of codewords, the GCD value is used to find correlation for different candidate polynomials. For valid block length and correct generator polynomial, this correlation accumulates to a very high value. For some pair of codewords the generator polynomial is not exposed, however by factorizing the detected polynomial under modulo-2 operation, the desired generator polynomial is retrieved and the correlation value increases further. Upon plotting the correlation values, the desired generator polynomial for valid block length is exposed very explicitly.

The proposed technique exploits the cyclic relationship between codewords of BCH codes. This technique works perfectly for noiseless bitstream. However, it is equally valid if there are certain errors in some of the codewords. Since this is an analysis technique unlike realtime decoding, the effects of noisy codewords can be reduced by increasing the number of codewords. The correlation value accumulates for increased number of test vectors with very mild increase in processing and hence the algorithm works for noisy bitstream as well.

This paper is organized as follows. In Section II, we recall the principles of BCH code construction along with standard procedure to generate test vectors. Section III gives a refresher about GCD and Euclid's algorithm following the detection of generator polynomial mathematics in Section IV. Simulation results are shown in Section V.

\section{BCH construction}\label{sec:bch_construction}
Given any positive integer $m$ $(m\geq3)$ and error correcting capability $t$ $(t<2^{m-1})$, a BCH code can be generated with the following parameters:-

\begin{tabbing}
Number of parity-check digits: \= $n-k \leq mt,$ \kill
  Block Length: \>  $n=2^{m-1},$ \\
  Minimum distance: \> $d_{min}\geq 2t+1,$ \\
  Number of parity-check digits: \> $n-k \leq mt,$
\end{tabbing}

The generator polynomial $g(X)$ is the LCM of $\phi_1(X), \phi_2(X), \cdot\cdot\cdot, \phi_{2t}(X):$
\begin{equation}\label{equation:eq1}
    g(X) = LCM\{\phi_1(X), \phi_2(X), \cdot\cdot\cdot, \phi_{2t}(X)\}
\end{equation}

Since every even power of primitive element $\alpha$ has the same minimal polynomial as some preceding odd power of $\alpha,$ hence (\ref{equation:eq1}) can be reduced to:-
\begin{equation}\label{equation:eq2}
    g(X) = LCM\{\phi_1(X), \phi_3(X), \cdot\cdot\cdot, \phi_{2t-1}(X)\}
\end{equation}

Test vectors (encoding in systematic form) are generated by following the standard encoding steps \cite{ShuLin} which are:-
\begin{enumerate}
  \item Pre-multiply, $k$ information digits, message polynomial with $X^{n-k}$ i.e. $X^{n-k} u(X)$.
  \item Calculate parity check polynomial $b(X)$ from dividing $X^{n-k} u(X)$ by $g(X)$.
  \item Append $b(X)$ with $X^{n-k} u(X)$ to obtain the code polynomial $v(X)= b(X)+ X^{n-k} u(X)$.
\end{enumerate}
	
\begin{figure}[!ht]
  \includegraphics[width=0.96\columnwidth]{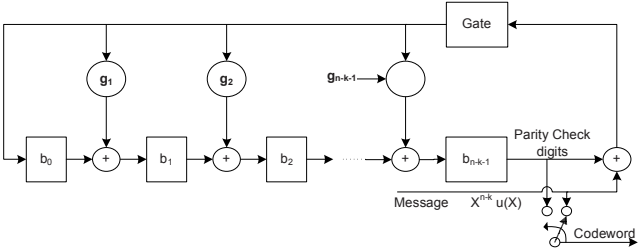}
  \caption{Encoding circuit for an $(n,k)$ cyclic code with generator polynomial
    \quad $g(X)= 1+g_1X+g_2X^2+\cdot\cdot\cdot+g_{n-k-1}X^{n-k-1}+X^{n-k}$}\label{figure:fig1}
\end{figure}

The above steps can be realized by a division circuit based on linear $(n-k)$ stage shift register with feedback connections based on $g(X)$ as shown in Figure \ref{figure:fig1}.

The operation of the encoding circuit \cite{ShuLin} is described as follows:
\begin{enumerate}
  \item Initially, the gate is turned on. The, $k$ information digits, message polynomial $u(X)= u_0+u_1X+\cdot\cdot\cdot+u_{k-1}X^{k-1}$ is fed to the circuit as well as transmitted into the channel. Feeding the $k$ information digits into the circuit is equivalent to pre-multiplying $u(X)$ by $X^{n-k}$. When all $k$ information digits are shifted into the circuit, the $(n-k)$ digits in the register form the remainder.
  \item The gate is then turned off, since the register now contains the desired $(n-k)$ parity check digits.
  \item Selector is changed to the right position to send parity check digits into the channel. These $(n-k)$ parity check digits along with $k$ information digits form the cyclic codeword in systematic form.
\end{enumerate}

\section{Greatest Common Divisor}\label{sec:GCD}
To have insight into the detection algorithm, some basic definitions \cite{Stephen} are described as follows:-
\begin{itemize}
  \item Common Divisor:	An element $a$ is a common divisor of a collection of elements ${b_1,b_2,\cdot\cdot\cdot,b_n}$ if $a$ divides all elements of $b_i$ for $i=1,2,\cdot\cdot\cdot,n$ with remainder zero.
  \item Greatest Common Divisor:	If $d$ is a common divisor of ${b_i}$ and all other common divisors are less than $d$, then $d$ is called the greatest common divisor (GCD) of the ${b_i}$.
\end{itemize}

Euclid's algorithm \cite{Stephen} is very famous for fast GCD calculations. This algorithm is outlined as follows:-
\begin{enumerate}
  \item Let $a, b$ be two elements where $a>b$.
  \item Let $r_i$ take on initial value $r_{-1}=a$ and $r_0=b$.
  \item If $r_{i-1}\neq 0,$ then define $r_i$ using $r_{i-2}+q_ir_{i-1}= r_i,$ where $r_i<r_{i-1}$.
  \item If $r_i=0,$ then $r_{i-1}= GCD(a,b)$, else goto step 3.
\end{enumerate}

In our work, Euclid's algorithm over polynomials is used to find GCD among different code polynomials. The code polynomials are passed to the algorithm in a combinatorial manner to maximize the correlation value for desired generator polynomial.

\section{Detection of Generator Polynomial}\label{sec:Detection of g(x)}
For cyclic codes, every codeword in code space $C$ is obtained by polynomial multiplication of message and generator. So this algebraic structure can readily reveal the generator polynomial in code polynomials. For the sake of clarity, we prove the following:-

\textbf{Proposition}:
The GCD polynomial, $\textbf{d}_{c}(X)$ of any two code polynomials from code space $C$ contains generator polynomial, \textbf{g}(X) as one of its factor.

\textbf{Proof}:
Let GCD($\textbf{m}_{1}(X),\textbf{m}_{2}(X)$) = $\textbf{d}_{m}(X)$\\
where $1\leq \textbf{d}_{m}(X)\leq min(\textbf{m}_{1}(X),\textbf{m}_{2}(X))$ for any $\textbf{m}_{1}(X),\textbf{m}_{2}(X)\in M$\\

By definition of cyclic codes, we have $\textbf{c}_{1}(X) = \textbf{m}_{1}(X) g(X)$ and $\textbf{c}_{2}(X) = \textbf{m}_{2}(X) g(X)$\\
Therefore
\begin{eqnarray*}
    GCD(\textbf{c}_{1}(X),\textbf{c}_{2}(X))& = GCD(\textbf{m}_{1}(X) \textbf{g}(X),\textbf{m}_{2}(X) \textbf{g}(X))\\
       & = \textbf{g}(X). GCD(\textbf{m}_{1}(X),\textbf{m}_{2}(X))\\
       & = \textbf{g}(X).\textbf{d}_{m}(X)\\
       & = \textbf{d}_{c}(X)
\end{eqnarray*}

Hence the proof.\\

\textbf{Corollary}: If $\textbf{m}_{1}(X)$ and $\textbf{m}_{2}(X)$ are co-prime, then GCD($\textbf{m}_{1}(X), \textbf{m}_{2}(X)$) = 1
then cd = $\textbf{g}(X)$

For illustration purpose, $(7,4)$ cyclic code generated by $g(X)= 1+X+X^3$ is chosen. Some of the message vectors, code vectors and code polynomials are shown in Table \ref{Table:tab1}.

\begin{table}[!ht]
   \centering
   \caption{Code Polynomials for (7,4) cyclic code}\label{Table:tab1}
   \begin{tabular}{|c|c|l|}
   \hline
    Message & \multicolumn{1}{c|}{Code} & \multicolumn{1}{c|}{Code}\\
    Vector & \multicolumn{1}{c|}{Vector} & \multicolumn{1}{c|}{Polynomials}\\
   \hline
   1000 & 1101000 & $X^3+X+1=\bf g(X)$\\[6pt]
   \hline
   1010 & 0011010 & $X^5+X^3+X^2=X^2\bf g(X)$\\[6pt]
   \hline
   0110 & 1000110 & $X^5+X^4+1=(X^2+X+1)\bf g(X)$\\[6pt]
   \hline
   1110 & 0101110 & $X^5+X^4+X^3+X=(X^2+X)\bf g(X)$\\[6pt]
   \hline
   1001 & 0111001 & $X^6+X^3+X^2+X=(X^3+X)\bf g(X)$\\[6pt]
   \hline
   0111 & 0010111 & $X^6+X^5+X^4+X^2=(X^3+X^2)\bf g(X)$\\[6pt]
   \hline
   \end{tabular}
\end{table}

Each code polynomial of cyclic code carry the shift relationship, imparted by $g(X)$, which can be exploited by calculating the GCD for any two code polynomial. Let two non-zero noiseless dissimilar code polynomials $v_1(X)$ and $v_2(X)$ are transmitted. The received code polynomials are:-
\begin{equation}\label{equation:eq3}
    \begin{aligned}
    r_1(X)= v_1(X)+ e_1(X)\\
    r_2(X)= v_2(X)+ e_2(X)
    \end{aligned}
\end{equation}

Since code polynomials are assumed to be noise-free hence $e_1(X) = e_2(X) = 0$ and (\ref{equation:eq3}) are reduced to:-
\begin{equation}\label{equation:eq4}
    \begin{aligned}
	r_1(X)= v_1(X)\\
    r_2(X)= v_2(X)
    \end{aligned}
\end{equation}

The GCD calculation on these noise-free code polynomials results into detection of $g(X)$ either without factorization or with factorization under modulo-2 operation.

\subsection{No Factorization for $g(X)$} \label{sec:prim_g(x)}
Suppose $r_1(X)=X^2+X^4+X^5+X^6$ and $r_2(X)=1+X+X^3$ from Table \ref{Table:tab1} are received . Division operation will result in $a(X)=q(X)b(X)+r(X),$ where $a(X)$ and $b(X)$ are first and second code polynomials respectively, $q(X)$ is the quotient polynomial, $r(X)$ is the remainder polynomial and $+$ shows modulo-2 addition.
\begin{equation*}
    \begin{aligned}
       X^2+X^4+X^5+X^6 & = X^3.(1+X+X^3)\\
       & \quad + \underbrace{\bf(X^2+X^3+X^5)}\Rightarrow \bf r(X) \neq 0
     \end{aligned}
\end{equation*}

Carry on with Euclid's algorithm till $r(X)$ becomes zero.
\begin{equation*}
    \begin{aligned}
       X^2+X^3+X^5 & = X^2.(1+X+X^3)\\
       & \quad + \underbrace{\bf 0}\Rightarrow \bf r(X) = 0
     \end{aligned}
\end{equation*}

Since $r(X)$ is zero, hence $q(X)= 1+X+X^3$ is the greatest common divisor of $r_1(X)$ and $r_2(X)$. Hence GCD of two code polynomials directly results into $g(X)$.

\subsection{Factorization for $g(X)$} \label{sec:mult_g(x)}
Now suppose $r_1(X)=X+X^2+X^3+X^6$ and $r_2(X)=X^2+X^4+X^5+X^6$ from Table \ref{Table:tab1} are received. The GCD calculation will proceed as follows:
\begin{equation*}
    \begin{split}
       X+X^2+X^3+X^6 & =1.(X^2+X^4+X^5+X^6)\\
       & \quad + \underbrace{\bf(X+X^3+X^4+X^5)}\\
       & \qquad \qquad \qquad \Rightarrow\bf r(X) \neq 0
     \end{split}
\end{equation*}
Carry on with second iteration.
\begin{equation*}
    \begin{split}
       X^2+X^4+X^5+X^6 & =X.(X+X^3+X^4+X^5)\\
       & \quad + \underbrace{\bf 0}\Rightarrow \bf r(X) = 0
     \end{split}
\end{equation*}

Since $r(X)$ is zero, hence $q(X)= X+X^3+X^4+X^5$ is the greatest common divisor of $r_1(X)$ and $r_2(X)$. At first glance it looks very different from $g(X)$ but it can be reduced to $g(X)$ by factorization under modulo-2 operation.
\begin{equation*}
    \begin{split}
       X+X^3+X^4+X^5 & =\underbrace{\bf(X^3+X+1)}(X^2+X)\\
       &  \qquad \quad \bf g(X)
     \end{split}
\end{equation*}
Hence GCD exploited the cyclic shift relation between codewords along with factorization (if needed) under modulo-2 operation to detect generator polynomial.

\begin{figure*}
    \centering
    \subfigure[$(31,26,1)$ BCH code]
    {
        \includegraphics[width = 0.96\columnwidth, height = 2in]{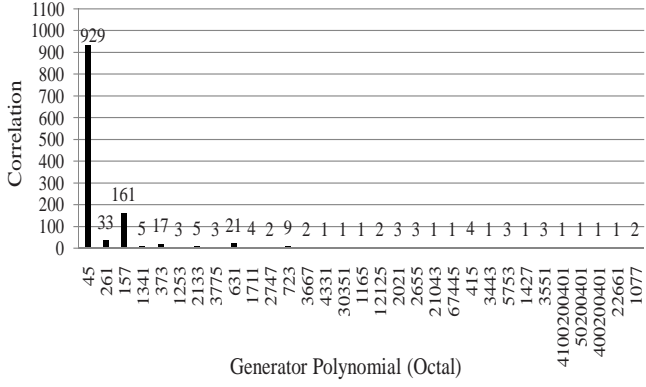}
        \label{fig:first_sub}
    }
    \subfigure[$(31,21,2)$ BCH code]
    {
        \includegraphics[width = 0.96\columnwidth,height = 2in]{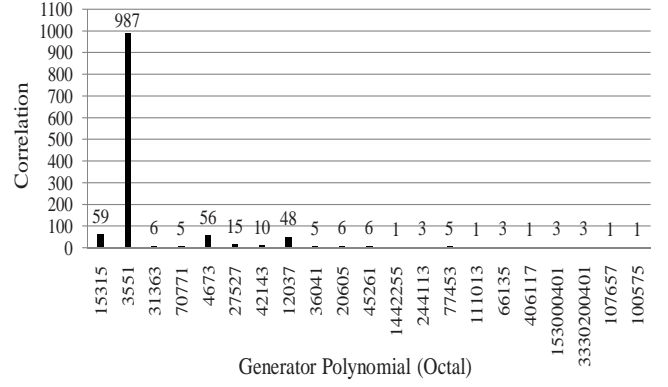}
       \label{fig:second_sub}
    }

    \subfigure[$(31,16,3)$ BCH code]
    {
        \includegraphics[width = 0.96\columnwidth, height = 2in]{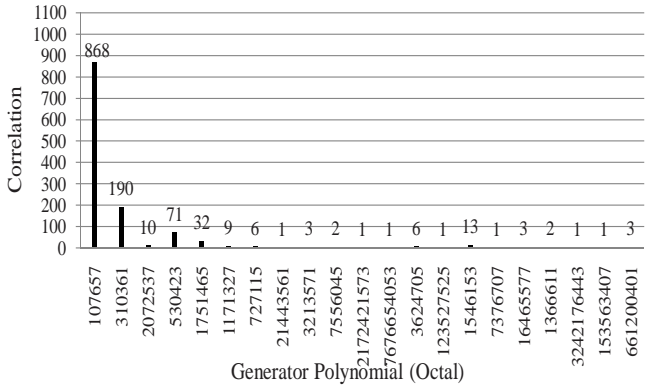}
        \label{fig:third_sub}
    }
     \subfigure[$(31,11,5)$ BCH code]
    {
        \includegraphics[width = 0.96\columnwidth, height = 2in]{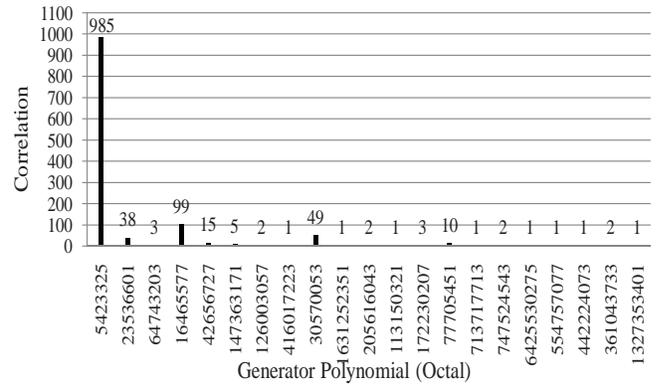}
        \label{fig:third_sub}
    }
    \caption{Simulation results for $(31,k,t)$ BCH code.}
    \label{figure:fig2}
\end{figure*}

\section{Simulation Results}\label{sec:Simulation}
The algorithm is tested on a wide range of $(n,k,t)$ combinations of BCH codes. The code polynomials from the test vector are stepped through the algorithm in a combinatorial manner. The GCD of first code polynomial is calculated with all other code polynomials in a descending order. Then GCD of second code polynomial is calculated with all other code polynomials in a descending order. This process is continued till GCD calculation of last two candidate code polynomials. The resulting polynomials are correlated and the correlation value for the generator polynomial of test vectors is found to be very high. For illustration purpose, fifty BCH codewords, encoded by $(31,k,t)$ parameter, are chosen. These code polynomial are given to the algorithm for two different scenarios:-
\subsection{Known Block Length} \label{sec:KnownBL}
In first case, by fixing the block length, first fifty code polynomials are passed to the algorithm. The simulation results for $(31,26,1), (31,21,2), (31,16,3)$ and $(31,11,5)$ codes are shown in Figure \ref{figure:fig2}. In these figures, polynomial (octal form) are plotted on horizontal axis and corresponding correlation values are plotted on vertical axis.

In  Figure \ref{figure:fig2} \{a, b, c \& d\}, correlation values of 929, 987, 868 and 985 for polynomials $p(X)= 45$, $p(X)= 3551$, $p(X)= 107657$ and $p(X)= 5423325$ are shown respectively. These $p(X)$ corresponds to generator polynomial $g(X)$. The above correlation values correspond to fifty noiseless code polynomials. This value depends on number of code polynomials chosen and the noise present in code polynomials. In case of noisy code polynomials, this value can be smaller and it can possibly be increased by increasing the number of code polynomials for GCD calculation. Correlation value found for $p(X)= g(X)$ in different simulations is reasonably high as compared to all other polynomials.

\subsubsection{Competitive Polynomial Analysis}
The competitive correlation values (Figure \ref{figure:fig2}(a)) for octal polynomials 157, 261, 631, 373, 723, 1341, 1711, 1253, 2747 and 4331 can be 161, 33, 21, 17, 9, 5, 4, 3, 2 and 1  respectively. For illustrative purpose only unique correlation values and corresponding polynomials are chosen for analysis. If the chosen polynomials are factored under modulo-2 operation, they result into the desired generator polynomial. This can be shown as follows:-

\begin{eqnarray*}
    p(X) & = 157(oct)\\
       & = X^6+ X^5+ X^3+ X^2+ X+ 1\\
       & = (X+1) \underbrace{\bf (X^5+X^2+1)}\\
    p(X) & = 261(oct)\\
       & = X^7+ X^5+ X^4+ 1\\
       & = (X+1)^2 \underbrace{\bf (X^5+X^2+1)}\\
    p(X) & = 631(oct)\\
       & = X^8+ X^7+ X^4+X^3+1\\
       & = (X^3+X^2+1) \underbrace{\bf (X^5+X^2+1)}\\
    p(X) & = 373(oct)\\
       & = X^7+ X^6+ X^5+X^4+X^3+X+1\\
       & = (X^2+X+1) \underbrace{\bf (X^5+X^2+1)}\\
    p(X) & = 723(oct)\\
       & = X^8+X^7+X^6+ X^4+X+1\\
       & = (X+1)^3 \underbrace{\bf (X^5+X^2+1)}\\
    p(X) & = 1341(oct)\\
       & = X^9+ X^7+ X^6+ X^5+ 1\\
       & = (X^2+X+1)^2 \underbrace{\bf (X^5+X^2+1)}\\
    p(X) & = 1711(oct)\\
       & = X^9+X^8+ X^7+X^6+X^3+1\\
       & = (X+1)(X^3+X+1) \underbrace{\bf (X^5+X^2+1)}\\
    p(X) & = 1253(oct)\\
       & = X^9+ X^7+ X^5+X^3+X+1\\
       & = (X^3+X^2+1) \underbrace{\bf (X^5+X^2+1)}\\
    p(X) & = 2747(oct)\\
       & = X^{10}+X^8+ X^7+X^6+X^5+X^2+X+1\\
       & = (X+1)(X^4+X^3+1) \underbrace{\bf (X^5+X^2+1)}\\
    p(X) & = 4331(oct)\\
       & = X^{11}+X^7+X^6+ X^4+X^3+1\\
       & = (X+1)(X^5+X^4+X^3+X+1) \underbrace{\bf (X^5+X^2+1)}
\end{eqnarray*}

Although $g(X)= 4100200401$(oct) seems to be very complicated but it gets factored and one part is again the desired polynomial.
\begin{equation*}
    \begin{split}
       p(X) & = 4100200401(oct)\\
       & = X^{29}+X^{24}+X^{16}+ X^8+1\\
       & = (X^{12}+X^{11}+X^9+X^5+1)\\
       &\quad (X^9+X^7+X^6+X^5+X^4+X^3+1)\\
       &\quad \bf (X^5+X^2+1)\Longrightarrow \bf g(X)\\
       &\quad (X^3+X^2+1)
     \end{split}
\end{equation*}

\subsubsection{Correlation for Incorrect block length}
\begin{figure}[!htb]
  \includegraphics[width=0.96\columnwidth,height = 2in]{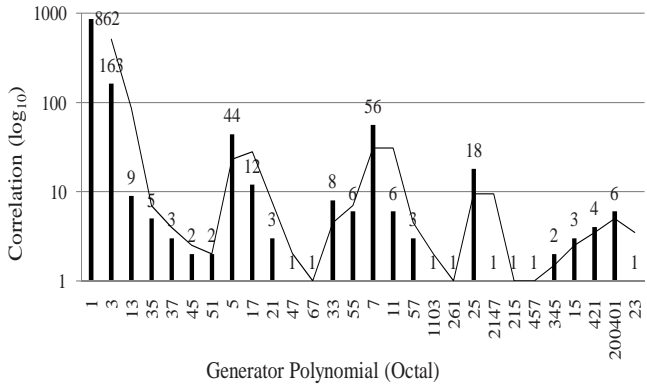}
  \caption{Correlation trend for incorrect block length}\label{figure:fig3}
\end{figure}

If the block length is incorrect, a very intuitive correlation trend can be seen. It is obvious that $g(X)= 1 (oct)= 1$ is a factor of every higher order polynomial, so its correlation value has to be higher than all other polynomials. Similarly irreducible polynomials $(1+X),(1+X^2 ),(1+X+X^2)$ etc can be factors of some higher order polynomials and hence they will show higher correlation values as compared to other candidates. This trend can be seen in Figure \ref{figure:fig3}.

\subsection{Unknown Block Length} \label{sec:UnknownBL}
\begin{figure}[!htb]
  \includegraphics[width=0.96\columnwidth,height = 2in]{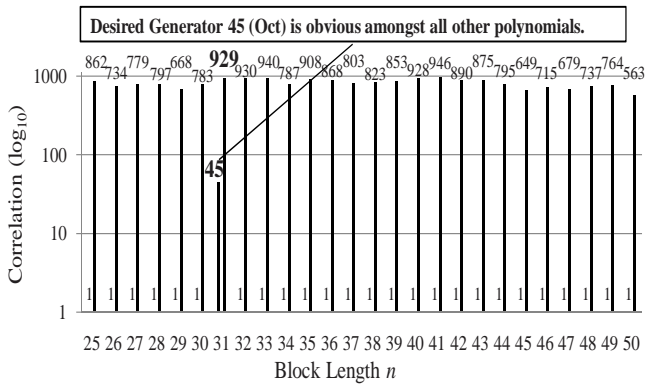}
  \caption{Correlation results for variable block length}\label{figure:fig4}
\end{figure}

The second scenario is simulated for $g(X)= X^5+X^2+1=45(oct)$ with a block length of 31. The algorithm is run by varying the block length for a range of values e.g. $n=25$ to $50$. Here maximum correlation (close to the desired polynomial) is found for $g(X)= 1$  as compliant to correlation trend shown in Figure \ref{figure:fig3}. It is obvious that $g(X)= 1$ cannot be a generator polynomial of binary primitive BCH code as it does not meet $n-k \leq mt$ criterion. In Figure \ref{figure:fig4}, correlation value bar is plotted along with corresponding polynomial bar. Here the polynomial bar is invisible for incorrect block lengths, however polynomial bar along with corresponding correlation bar is high enough to show that the correct block length is 31 with a generator polynomial $g(X)= X^5+X^2+1=45(oct)$.

\section{Conclusion}
The proposed algorithm exploits the cyclic relationship between code polynomials of BCH codes. It calculates greatest common divisor between different received code polynomials in a combinatorial manner and finds the corresponding maximum correlation. It takes into account two possible scenarios of known and unknown prior knowledge of block length. The simulation results show that the correlation for the noiseless code polynomials is very high as compared to other candidate polynomials which are in fact not the competitive polynomials but they are some higher order multiples of generator polynomial. These high order multiples can be reduced to generator polynomial by factorization under modulo-2 operations.

In simulation, only noiseless codewords are used, however intuitively it can work on noisy codewords as well. The only minute difference will be reduction in a correlation value due to noise effects. This reduction in correlation value can be taken care of by passing large number of codewords to the algorithm.


\bibliographystyle{IEEEtran}
\bibliography{Det_BCH_Gen}

\end{document}